\documentclass[10pt,aps,prd,twocolumn,superscriptaddress,nofootinbib]{revtex4-2}
\newcommand{\ibar}{{\declareslashed{}{\text{-}}{0.04}{0}{I}\slashed{I}}}
\newcommand{\jcap}{J. Cosmol. Astropart. Phys. }

\newcommand{\apjl}{Astrophys. J. Lett.}
\newcommand{\physrep}{Phys. Rep. }
\newcommand{\aap}{Astron. Astrophys. }

\newcommand{\mnras}{Mon. Not. R. Astron. Soc. }

\newcommand{\nphysa}{Nucl. Phys. A }

\usepackage{color}
\usepackage{amsmath}
\usepackage{graphicx}
\usepackage{slashed}
\usepackage{hyperref}
\hypersetup{
    colorlinks=true,
    linkcolor=blue,
    filecolor=magenta,      
    urlcolor=cyan,
    citecolor=blue
}

\begin{document}

\title{Core-collapse supernova explosions hindered by eV-mass sterile neutrinos}

\author{Kanji Mori}
\email[]{kanji.mori@nao.ac.jp}
\altaffiliation{Research Fellow of Japan Society for the Promotion of Science}
\affiliation{National Astronomical Observatory of Japan, 2-21-1 Osawa, Mitaka, Tokyo 181-8588, Japan}
\author{Tomoya Takiwaki}
\affiliation{National Astronomical Observatory of Japan, 2-21-1 Osawa, Mitaka, Tokyo 181-8588, Japan}
\author{Kazunori Kohri}
\affiliation{National Astronomical Observatory of Japan, 2-21-1 Osawa, Mitaka, Tokyo 181-8588, Japan}
\affiliation{Institute of Particle and Nuclear Studies, and International Center for Quantum-field Measurement Systems for Studies of the Universe and Particles (WPI), KEK, 1-1 Oho, Tsukuba, Ibaraki 305-0801, Japan}
\affiliation{Kavli Institute for the Physics and Mathematics of the Universe (WPI), The University of Tokyo, 5-1-5 Kashiwanoha, Kashiwa, Chiba 277-8583, Japan}
\author{Hiroki Nagakura}
\affiliation{National Astronomical Observatory of Japan, 2-21-1 Osawa, Mitaka, Tokyo 181-8588, Japan}

\date{\today}

\begin{abstract}

Light sterile neutrinos, $\nu_s$, are often introduced to explain an anomalous deficit in the electron antineutrino flux from nuclear reactors. If they exist, sterile neutrinos would also be produced in collapsing massive stars through the active-sterile neutrino oscillation. In order to investigate the impacts of sterile neutrinos on supernova dynamics, we perform two-dimensional neutrino-radiation hydrodynamic simulations of stellar core-collapse coupled with the active-sterile oscillation through the Mikheyev–Smirnov–Wolfenstein effect.  As the initial condition of our simulations, we adopt a blue supergiant model  that is tuned to reproduce observational features of the SN~1987A progenitor to compare our models with observations of the event. It is found that the active-sterile oscillation reduces the $\nu_{e}$ and $\bar{\nu}_e$ fluxes and decreases the explosion energy. We also find that, if the mixing angle $\theta$ and the mass difference $\delta m_\mathrm{s}^2$ between $\nu_e$ and $\nu_s$ are large enough, the star fails to explode. This suggests that these mixing parameters relevant to sterile neutrinos could be constrained by supernova explodability, though other uncertainties in supernova theory need to be addressed to refine them. In addition, we predict neutrino signals from a nearby supernova event and find that the neutrino event number can significantly decrease because the $\nu_e$ and $\bar{\nu}_e$ fluxes are reduced. In particular, DUNE observations of $\nu_e$ will be useful to search for a signature of sterile neutrinos with a tiny mixing angle because a smaller mixing angle leads to a larger effect on the $\nu_e$ flux.
\end{abstract}

\maketitle

\section{Introduction}
Neutrinos from nuclear reactors have been measured to reveal the existence and the nature of the particle \cite{1956Sci...124..103C,1956Natur.178..446R}. Such reactor experiments have precisely determined the neutrino oscillation parameters. However, there is a discrepancy between the observed flux of reactor electron antineutrinos \cite{2003EPJC...27..331A,2008PhRvL.100v1803A,2014JHEP...10..086A,2016PhRvL.116u1801C,2017PhRvL.118l1802K,2018PhRvL.121x1805A,2019JETPL.109..213S,2021PhRvD.103c2001A,2023Natur.613..257S} and recent theoretical predictions \cite{2011PhRvC..83e4615M,2011PhRvC..84b4617H}. This reactor antineutrino anomaly (RAA) could be attributed to the oscillation of electron antineutrinos into eV-mass sterile neutrinos \cite{2011PhRvD..83g3006M,2018JHEP...08..010D,2021JHEP...01..167B,2021PhR...928....1D}, although uncertainties in the reactor flux modeling are still under debate as well \cite{2014PhRvL.112t2501H,2019PhRvL.123b2502E,2022PhLB..82937054G,2023PhRvL.130b1801L}. Apart from the RAA, there is an unexplained bump at 5\,MeV in the reactor neutrino spectrum \cite{2014JHEP...10..086A,2015PhRvL.114a2502D,2017PhRvL.118d2502H,2019PhRvD..99e5045B}. In addition,  measurements of neutrinos from radioactive sources with gallium targets show a deficit of the event number compared with predictions \cite{2006PhRvC..73d5805A,2010PhLB..685...47K}.

Theoretically, right-handed sterile neutrinos are often introduced through the seesaw mechanism to explain the smallness of the active neutrino mass. Although the simplest type I seesaw mechanism requires superheavy sterile neutrinos \cite{1977PhLB...67..421M,Gell-Mann:1979vob,1980PhRvL..44..912M,1980PThPh..64.1103Y}, there are variants of the mechanism such as the neutrino minimal standard model \cite{2005PhLB..631..151A,2011JHEP...07..091B} and the inverse seesaw model \cite{1986PhRvL..56..561M}, in which we can obtain light sterile neutrinos.

The experimental anomalies and the theoretical models motivate astrophysical searches of light sterile neutrinos. A major astrophysical source of neutrinos is the Sun. Measurements of Solar neutrinos \cite{2008PhRvL.101k1301A,2011PhRvL.107n1302B,2016PhRvD..94e2010A} provide an upper limit $\sin^22\theta\lesssim0.3$ \cite{2018JHEP...08..010D}, where $\theta$ is the mixing angle between sterile and electron neutrinos. Also, some parameters in the $m_s$--$\sin^2 2\theta$ plane might have a tension with cosmological observations such as Cosmic Microwave Background \cite[e.g.,][]{Gariazzo2016}
 or Big Bang Nucleosynthesis. However, in some scenarios motivated by a new theory beyond the standard model such as supersymmetry or supergravity, the tension is relieved {\cite{2020JCAP...08..015H}}.

Another possible source of sterile neutrinos is the core-collapse of massive stars. Since sterile neutrinos are not involved in the weak interaction, the initial condition for the neutrino propagation is the flavor eigenstates of active neutrinos. However, sterile states can appear through the Mikheyev-Smirnov-Wolfenstein (MSW) effect \cite{1979PhRvD..20.2634W,1986NCimC...9...17M,PhysRevLett.56.1305} during the propagation in the stellar mantle. The production of eV-mass sterile neutrinos in core-collapse supernovae has been discussed in Refs.~\cite{1991NuPhB.358..435K,1992A&A...254..121P,1993APh.....1..165R,1997PhRvD..56.1704N,1999PhRvC..59.2873M,2000PhRvD..61l3005C,2003APh....18..433F,2006PhRvD..73i3007B,2007PhRvD..76l5026K,2012JCAP...01..013T,2014PhRvD..89f1303W,2015ApJ...808..188P,2019ApJ...880...81X,2020JCAP...10..038T,2020ApJ...894...99K}. These studies revealed that light sterile neutrinos reduce active neutrino fluxes and suppress neutrino heating  significantly \cite{2012JCAP...01..013T,2014PhRvD..89f1303W}. In addition, if they mix with electron neutrinos, they reduce the electron fraction and thus enhance the synthesis of heavy elements beyond iron \cite{2014PhRvD..89f1303W,2015ApJ...808..188P,2019ApJ...880...81X,2020ApJ...894...99K}. However, the previous studies employed the post-process or a steady-state model \cite{1986ApJ...309..141D,1996ApJ...471..331Q} for the neutrino-driven wind to investigate the effects of sterile neutrinos. Although these methods can provide qualitative predictions, nonlinear feedback on dynamics can be fully discussed only by self-consistent multi-dimensional core-collapse simulations \cite[e.g.,][]{2014ApJ...785..123C,2014ApJ...786...83T,2015ApJ...807L..31L,2015ApJ...808L..42M,2015ApJ...801L..24M,2016ApJ...817...72P,2016ApJ...817...72P,2017MNRAS.472..491M,2019JPhG...46a4001P,2018ApJ...855L...3O,2020MNRAS.491.2715B,2020ApJ...896..102K,2021ApJ...915...28B,2022MNRAS.516.1752M}. In particular, the effects of exotic particles are discussed in Refs.~\cite{Betranhandy2022,2023PhRvD.108f3027M,2024PhRvD.110b3031M}.

The reduction of the active neutrino flux could lead to failures of supernova explosions. Since type II supernovae obviously explode in nature, the supernova explodability is expected to provide a constraint on sterile neutrinos. In order to obtain the new constraint, performing self-consistent simulations are necessary. However, since even modern supernova simulations are subject to large uncertainties, it is difficult to establish the constraint in their current state. Still, future studies with more sophisticated simulations with reduced uncertainties would provide a reliable constraint. In this study, we aim to pave the way to establish the sterile neutrino constraint based on supernova explodability by performing self-consistent simulations with light sterile neutrinos for the first time.

Following this context, we develop two-dimensional models for a core-collapsing star coupled with the oscillation between sterile and active neutrinos for the first time. In particular, we assume sterile neutrinos which mix with electron (anti)neutrinos with the mixing angle $\sin^22\theta\sim0.04$ and the mass-squared difference $\delta m_\mathrm{s}^2\sim1$\,eV$^2$, which are motivated by the experimental hints from reactors. 

This paper is organized as follows. In Section~\ref{sec:method}, we explain our treatment of the neutrino oscillation and the setup of the simulations. In Section~\ref{sec:results}, we present the results of our simulations and predict observable quantities including the explosion energy and neutrino and gravitational wave (GW) signals. In Section~\ref{sec:sum-dis}, we discuss the implications of our result and future prospects.

 \section{Method}
 \label{sec:method}
\subsection{Neutrino Oscillation}
Neutrino oscillations can be described by the Shr\"{o}dinger-like equation \cite[e.g.][]{2019ApJ...880...81X}
\begin{equation}
    i\frac{d}{dr}\vec{\psi}(E,\;r)=H(E,\;r)\vec{\psi}(E,\;r),
\end{equation}
where $\vec{\psi}(E,\;r)$ is the neutrino wave function, $H(E,\,r)$ is the propagation Hamiltonian, $E$ is the neutrino energy, and $r$ is the radius. However, coupling the equation with hydrodynamic simulations is computationally expensive. Instead, we adopt the Landau-Zener formula \cite{1986PhRvL..57.1275P,1987PhRvD..35.4014K,1997PhRvD..56.1704N,2014PhRvD..89f1303W,2020JCAP...08..018S}, which can approximate the MSW effect.

The MSW resonance occurs when the condition
\begin{equation}
    \frac{\delta m_\mathrm{s}^2}{2E}\cos2\theta=V_\mathrm{eff} \label{res}
\end{equation}
is satisfied, where $V_\mathrm{eff}$ is the effective potential of the electron (anti)neutrino forward scattering on neutrons, protons, electrons, and positrons. The potential is given as \cite{1997PhRvD..56.1704N}
\begin{equation}
V_\mathrm{eff}=\pm\sqrt{2}G_\mathrm{F}n_\mathrm{b}\left(Y_e-\frac{1}{2}Y_n\right)\approx\pm\frac{3\sqrt{2}}{2}G_\mathrm{F}n_\mathrm{b}\left(Y_e-\frac{1}{3}\right)\label{Veff},
\end{equation}
where the plus sign is for $\nu_e$, the minus sign is for $\bar{\nu}_e$, $G_\mathrm{F}$ is the Fermi coupling constant,  $n_\mathrm{b}$ is the baryon number density, $Y_e$ is the electron mole fraction, and $Y_n$ is the neutron fraction. The condition~\eqref{res} for $\nu_e$ is satisfied when $Y_e\sim1/3$ or the baryon density is sufficiently low. Since the former condition is satisfied at a smaller radius in stars than the latter is, the resonance with $Y_e\sim1/3$ is called the inner resonance (IR) and the other is called the outer resonance (OR) \cite{2014PhRvD..89f1303W}. We checked with post-processing  that the IR typically appears at $r\sim20$--100\,km during the stellar core-collapse, while the OR appears at $r\gtrsim10^3$\,km. We hence only consider the IR in our simulations because the OR is located at the outside of the stalled shock. On the other hand, only the IR can happen for $\bar{\nu}_e$.

Apart from the MSW effect, the neutrino forward scattering on other neutrinos can also affect the active-neutrino oscillation. Since Refs.~\cite{2012JCAP...01..013T,2019ApJ...880...81X} report that the effect is small in the accretion phase at the post-bounce time $t_\mathrm{pb}<1$\,s, we do not take the collective oscillation into account in this study.

The conversion probability of electron (anti)neutrinos when they cross the resonance can be approximated by the Landau-Zener formula
\begin{equation}
    P_\mathrm{es}(E)=1-e^{-\frac{\pi^2}{2}\gamma(E)},\label{LZ}
\end{equation}
where $\gamma(E)=\Delta_\mathrm{IR}/l_\mathrm{osc}(E)$ is the adiabatic index, $l_\mathrm{osc}(E)=2\pi E/\delta m_\mathrm{s}^2\sin2\theta$ is the oscillation length, and
\begin{equation}
\Delta_\mathrm{IR}=\left|\frac{\frac{dV_\mathrm{eff}}{dr}}{V_\mathrm{eff}}\right|^{-1}\tan2\theta=\left|\frac{dV_\mathrm{eff}}{dr}\right|^{-1}\frac{\delta m_\mathrm{s}^2}{2E}\sin2\theta \label{delta}
\end{equation}
is the IR width. When $Y_e$ changes gradually in space, the resonance tends to be adiabatic, resulting in a complete flavor swap. As we shall show below, on the other hand, non-adiabaticity is not negligible for IR in some models. 

\begin{table*}[]
\begin{tabular}{cccccccccccc}
Model&Dimension&$\delta m_\mathrm{s}^2$  & $\sin^22\theta$&$\delta m_\mathrm{s}^2$ $\sin2\theta$ & Shock Revival? & $t_{\mathrm{pb},\,400}$ &$t_{\mathrm{pb},\,2000}$&$E_\mathrm{diag}$ & $M_\mathrm{Ni}$ & $M_\mathrm{PNS}$ \\
&&[eV$^2$] & &[eV$^2$]&&[ms] &[ms] &[$10^{51}$\,erg] &[$M_\odot$]&[$M_\odot$]\\

\hline
   \texttt{NoSterile}&2&--   & -- & -- &Yes&263 & 428& 0.28&0.062 & 1.73\\
\texttt{A}&2&3.90   & 0.040        &0.78&No&--&--&--&-- &-- \\
\texttt{B}&2&1.30   & 0.120        & 0.45&No&--&--&--&-- &-- \\

\texttt{C}&2&1.30   & 0.040        & 0.26 &Yes&442&610&0.12&0.016 &1.81   \\
\texttt{D}&2&1.30   & 0.013        & 0.15 &Yes&322&444&0.23&0.039&1.77  \\
\texttt{E}&2&0.43& 0.040        & 0.09 &Yes&312&452&0.26&0.048 &1.76  \\
\texttt{1D-C}&1&1.30& 0.040        & 0.26&No&--&--&--&-- &--  \\

\end{tabular}
\caption{The summary of our simulations. Model \texttt{NoSterile} is a reference model which does not consider sterile neutrinos, and Models \texttt{A}--\texttt{E} and \texttt{1D-C} consider sterile neutrinos with the mass-squared difference $\delta m_\mathrm{s}^2$ and the mixing angle $\sin^2 2\theta$. In Models \texttt{A} and \texttt{B}, the bounce shock is stalled until the end of the simulations, whereas  it is successfully revived in the other models. The bounce shock reaches $r=400$\,km at the post-bounce time $t_{\mathrm{pb}}=t_{\mathrm{pb},\;400}$, which is shown in the sixth column and indicates when the shock wave is revived.  We estimate the diagnostic explosion energy $E_\mathrm{diag}$, the ejected nickel mass $M_\mathrm{Ni}$, and the proto-neutron star mass $M_\mathrm{PNS}$ at $t_\mathrm{pb}=t_{\mathrm{pb},\,2000}$, at which the bounce shock reaches $r=2000$\,km. The proto-neutron star is defined as a region where the density is higher than $10^{11}$\,g\,cm$^{-3}$.}
\label{table}
\end{table*}

\subsection{Hydrodynamic Simulations}
In this study, we perform one- and two-dimensional simulations of stellar core-collapse with the \texttt{3DnSNe} code \cite{2016MNRAS.461L.112T}. The code treats the active neutrino transport  with the three-flavor isotropic diffusion source approximation (IDSA) \cite{2009ApJ...698.1174L,2014ApJ...786...83T,2018ApJ...853..170K}. The nuclear equation of state is from the result in Ref.~\cite{1991NuPhA.535..331L} with $K=220$\,MeV. We also solve the nuclear reaction network \cite{2014ApJ...782...91N} that includes 13 species to take the energy generation into account. The simulated region is 5000\,km around the center of the star and the spatial resolution is $n_r=512$ for the one-dimensional models and $n_r\times n_\theta=512\times128$ for the two-dimensional models. We first perform one-dimensional simulations from the collapsing phase until $t_\mathrm{pb}=0.01$\,s and then switch them to two-dimensional geometry. For the two-dimensional simulations, the active-sterile oscillation is switched on at $t_\mathrm{pb}=0.01$\,s.

\begin{figure}
  \centering
  \includegraphics[width=7cm]{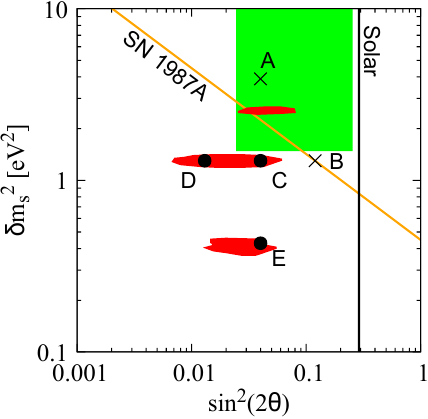}
  \caption{The $\delta m_\mathrm{s}$-$\sin^2(2\theta)$ parameter space. The two crosses represent the models without shock revival and the three circles represent the models with shock revival. The red regions are  the 99\% preferred parameters obtained from $\nu_e$ and $\bar{\nu}_e$ disappearance experiments \cite{2018JHEP...08..010D}  and the green region is the  parameters preferred to solve the RAA \cite{2011PhRvD..83g3006M}. The vertical line shows the upper limit on the mixing angle obtained from Solar neutrinos \cite{2018JHEP...08..010D}. The orange line indicates the SN~1987A explosion condition discussed in Section~\ref{sec:co-su-exp}.}
  \label{param}
 \end{figure}

We develop seven models with and without sterile neutrinos. The two-dimensional models with sterile neutrinos, Models \texttt{A}--\texttt{E}, are shown in Fig.~\ref{param} on the parameter space. The parameters we adopt are below an upper limit obtained from Solar neutrinos and are located around regions preferred by $\nu_e$ and $\bar{\nu}_e$ disappearance experiments~\cite{2018JHEP...08..010D}. In particular, Model \texttt{C} is located at the experimental best-fitting parameter derived in Ref.~\cite{2018JHEP...08..010D}. We call the model without sterile neutrinos Model \texttt{NoSterile}. Also, we develop a one-dimensional model with sterile neutrinos, Model \texttt{1D-C}, to investigate their effect on the collapsing phase.

 We adopt a $18.29M_\odot$ blue supergiant model in Ref.~\cite{2018MNRAS.473L.101U} as the initial condition of the simulation. This progenitor model is tuned to reproduce observational features of the SN~1987A progenitor Sk--69$^\circ$202, including its mass, lifetime, and position on the Hertzsprung-Russel diagram, on the basis of the slow-merger scenario \cite{1992ApJ...391..246P} of $14M_\odot$ and $9.0M_\odot$ stars. Three-dimensional hydrodynamic simulations of a supernova explosion with binary-merger progenitor models \cite{2017MNRAS.469.4649M,2018MNRAS.473L.101U} fine-tuned for SN~1987A have been performed in Refs.~\cite{2020ApJ...888..111O,2021ApJ...914....4U,2022MNRAS.514.3941N} and the consequences have been discussed in detail.

We consider the effect of the  oscillation between sterile and electron (anti)neutrinos during the simulations as follows. At each time step, the IR radii $r_\mathrm{IR}(E)$ for $\nu_e$ and $\bar{\nu}_e$ are determined by the condition~\eqref{res}. We then determine the conversion probability $P_\mathrm{es}(E)$ with the Landau-Zener formula~\eqref{LZ}. 

The time step in our simulations is $\Delta t\sim10^{-8}$--$10^{-7}$\,s, whereas the oscillation timescale is estimated as $\tau_\mathrm{osc}(E)=4\pi E/\delta m_\mathrm{s}^2\sin2\theta\sim10^{-6}$\,s. Since the oscillation timescale is longer than the simulation time step, one would overestimate the sterile neutrino flux if the oscillation is treated instantaneously at each time step. In order to avoid this problem, we employ the  Bhatnagar-Gross-Krook prescription \cite{1954PhRv...94..511B,2024PhRvD.109h3013N}
\begin{eqnarray}
    \frac{\partial f}{\partial t}=-\frac{1}{\tau_\mathrm{osc}}(f-f^\mathrm{a}),\label{BGK}
\end{eqnarray}
to treat the oscillation as a relaxation process, where $f$ is the electron (anti)neutrino distribution, $f^\mathrm{a}(E)=[1-P_\mathrm{es}(E)]f^\ast(E,\;r_\mathrm{IR})$ is the asymptotic distribution, and $f^\ast(E,\;r_\mathrm{IR})$ is the distribution at the intermediate time step obtained as the output of IDSA. The differential equation~\eqref{BGK} is readily solved as
\begin{eqnarray}
    f_{n+1}(E,\;r_\mathrm{IR})=f^\mathrm{a}(E)+(f^\ast(E,\;r_\mathrm{IR})-f^\mathrm{a}(E))e^{-\frac{\Delta t}{\tau_\mathrm{osc}(E)}},\label{BGK2}
\end{eqnarray}
where $f_{n+1}(E,\;r_\mathrm{IR})$ is the distribution at $r=r_\mathrm{IR}$ and at the next time step.

In the framework of IDSA, the active neutrino distribution is decomposed into the streaming component $f^\mathrm{s}(E,\;r)$ and the trapped one $f^\mathrm{t}(E,\;r)$. We apply Eq.~\eqref{BGK2} on $f^\mathrm{s}$ to estimate the streaming component at the next time step. On the other hand, when the trapped neutrinos oscillate into sterile neutrinos, the electron lepton number would be lost. This feedback effect is treated as follows.  IDSA keeps track of the number and energy densities of electron (anti)neutrinos
\begin{eqnarray}
    Y_{\nu_e(\bar{\nu}_{e})}^\mathrm{t}(r)=\frac{1}{n_\mathrm{b}}\frac{4\pi}{(hc)^3}\int f_{\nu_e(\bar{\nu}_{e})}^\mathrm{t}(E,\;r)E^2dE\nonumber\\
    Z_{\nu_e(\bar{\nu}_{e})}^\mathrm{t}(r)=\frac{1}{n_\mathrm{b}}\frac{4\pi}{(hc)^3}\int f_{\nu_e(\bar{\nu}_{e})}^\mathrm{t}(E,\;r)E^3dE,
\end{eqnarray}
at each time step, where $n_\mathrm{b}$ is the baryon number density and $f_{\nu_e(\bar{\nu}_{e})}^\mathrm{t}$ is the distribution for trapped electron (anti)neutrinos. We first assume the Fermi-Dirac distribution $f_{\nu_e(\bar{\nu}_{e})}^\mathrm{t}=1/(e^{\beta(E-\mu)}+1)$ and reconstruct the parameters $\beta$ and $\mu$ from $Y_{\nu_e(\bar{\nu}_{e})}^\mathrm{t}(r_\mathrm{IR})$ and $Z_{\nu_e(\bar{\nu}_{e})}^\mathrm{t}(r_\mathrm{IR})$. We then estimate $f_{\nu_e(\bar{\nu}_{e})}^\mathrm{t}$  at the next time step with Eq.~\eqref{BGK2} and recalculate $Y_{\nu_e(\bar{\nu}_{e})}^\mathrm{t}(r_\mathrm{IR})$ and $Z_{\nu_e(\bar{\nu}_{e})}^\mathrm{t}(r_\mathrm{IR})$, which are fed back to IDSA. 

\section{Results}\label{sec:results}

 We perform six stellar core-collapse simulations in two-dimensional geometry with and without sterile neutrinos. In this Section, we discuss the effect of sterile neutrinos on the post-bounce phase on the basis of the two-dimensional models. We also investigate their effect on the collapsing phase with the one-dimensional model in Appendix~\ref{sec:copha}. The adopted parameters and the results are summarized in Table~\ref{table}.

 \begin{figure}
  \centering
  \includegraphics[width=8.5cm]{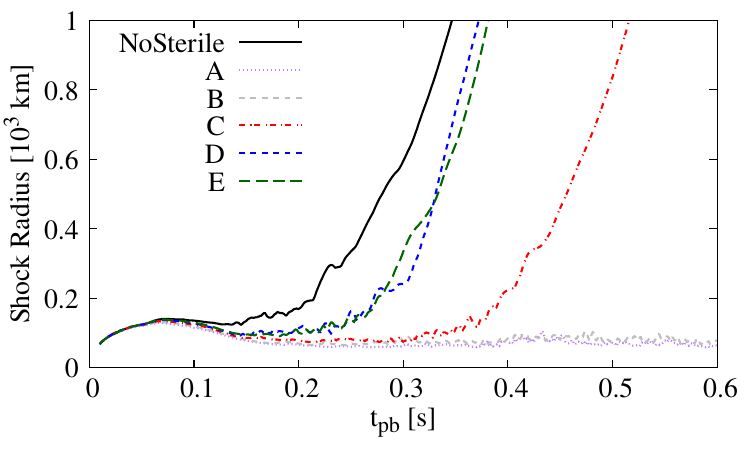}
  \caption{The angular-averaged radius of the bounce shock for each model as a function of the post-bounce time.}
  \label{rsh_2D}
 \end{figure}

 \subsection{Dynamics}

 Numerical simulations for stellar core-collapse have shown that the bounce shock stalls in the first $\mathcal{O}(100)$\,ms after the core bounce \cite[e.g.][]{1985ApJ...295...14B}. However, in the standard paradigm of the neutrino heating mechanism, the bounce shock can be revived later by the heating caused by the interaction between nucleons and active neutrinos. The introduction of sterile neutrinos could affect this picture. Since the active neutrino flux is reduced by the active-sterile oscillation, the shock revival could be delayed and even lead to the black hole formation. Fig.~\ref{rsh_2D} shows the shock radius for our models. When sterile neutrinos are not considered, the bounce shock is revived and the star successfully explodes. In Models \texttt{C}, \texttt{D}, and \texttt{E}, the shock revival is delayed compared with Model \texttt{NoSterile}, but the star still successfully explodes. In Models \texttt{A} and \texttt{B}, the shock is stalled until the end of the simulations. In this case, the explosion is failed and a black hole will be formed.
 
 \begin{figure}
  \centering
  \includegraphics[width=8.5cm]{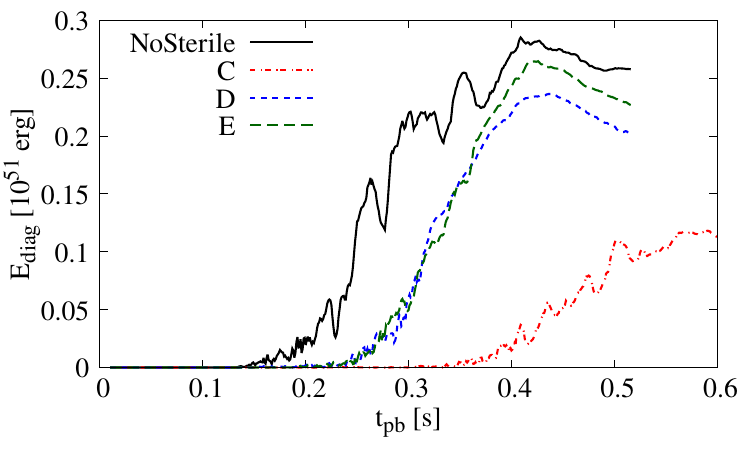}
  \caption{The diagnostic explosion energy for the exploding models as a function of the post-bounce time.}
  \label{Eexp}
 \end{figure}
 
 Figure \ref{Eexp} shows the diagnostic explosion energy
 \begin{eqnarray}
    E_\mathrm{diag}=\int_D dV\left(\frac{1}{2}\rho v^2+e-\rho\Phi\right),\label{Ediag}
\end{eqnarray}
where $\rho$ is the density, $v$ is the velocity, $e$ is the internal energy density, $\Phi$ is the gravitational energy potential, and $D$ is the region where the integrand is positive and the radial velocity is outward. Since the explosion energy is defined only for exploding models, Models \texttt{A} and \texttt{B} are not shown in Fig.~\ref{Eexp}. One can find that, for Model \texttt{NoSterile}, the explosion energy is as high as $0.25\times10^{50}$\,erg at the end of the simulation.  When sterile neutrinos are considered, the explosion energy becomes smaller than the one for Model \texttt{NoSterile}. 

Observationally, the explosion energy of SN~1987A is estimated to be $\sim(0.8$--$2.0)\times10^{51}$\,erg \cite{2014ApJ...783..125H,2020ApJ...888..111O}. The diagnostic energy of explosion in our models is significantly lower than the observed value. In general, multi-dimensional models tend to show $E_\mathrm{diag}$ smaller than $10^{51}$\,erg \cite{2019MNRAS.489..641M}. It is recently reported that the explosion energy in three-dimensional models continues to increase at $t_\mathrm{pb}>1$\,s and hence long-term simulations are necessary to predict the asymptotic value \cite{2021ApJ...915...28B,2024arXiv240106840B}. This implies that it is desirable to perform long-term simulations to impose a robust constraint on sterile neutrinos on the basis of the explosion energy.

\begin{figure}
  \centering

  \includegraphics[width=8.5cm]{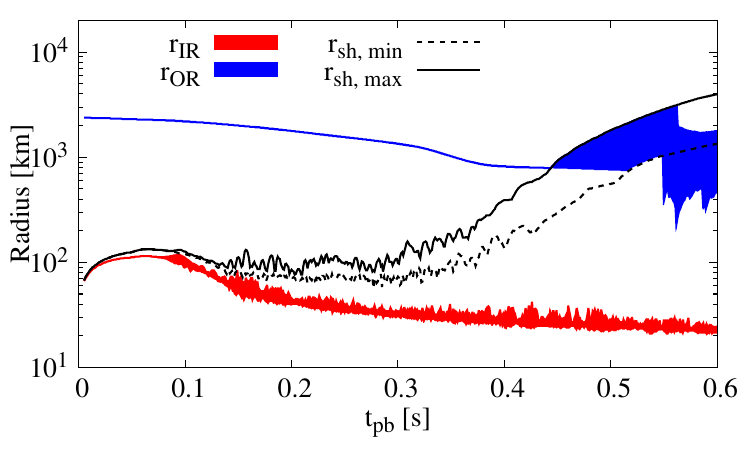}
  \caption{The IR and OR radii  for Model \texttt{C}. The bands indicate the range of the radii from minimum to maximum. The maximum and minimum shock radii are also shown.}
  \label{r_res}
 \end{figure}

 \begin{figure}
  \centering
  \includegraphics[width=8.5cm]{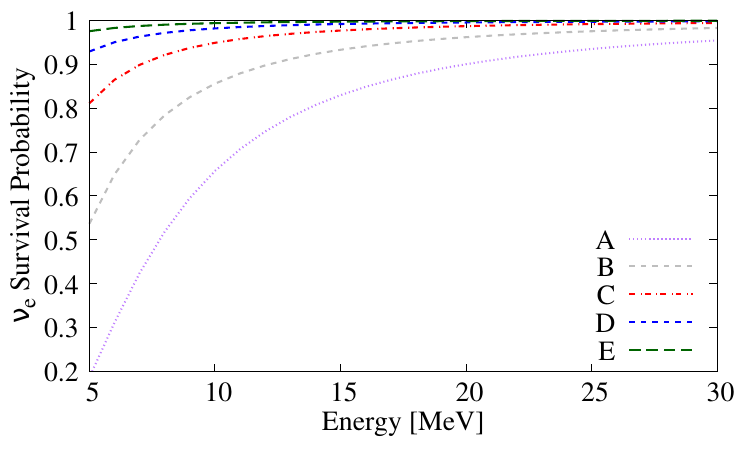}
  \caption{The angular-averaged $\nu_e$ survival probability $P_{ee}(E)$ evaluated at $t_\mathrm{pb}=0.05$\,s and $r=r_\mathrm{IR}\approx110$\,km as a function of the neutrino energy.}
  \label{P1}
 \end{figure}
  \begin{figure}
  \centering
  \includegraphics[width=8.5cm]{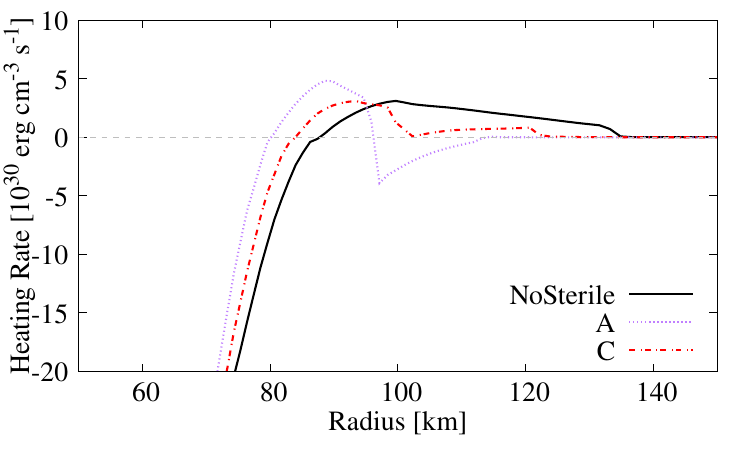}
  \caption{The angular-averaged profile for the neutrino heating rate at $t_\mathrm{pb}=0.1$\,s.}
  \label{Q}
 \end{figure}
 
 \begin{figure*}
  \centering
  \includegraphics[width=14cm]{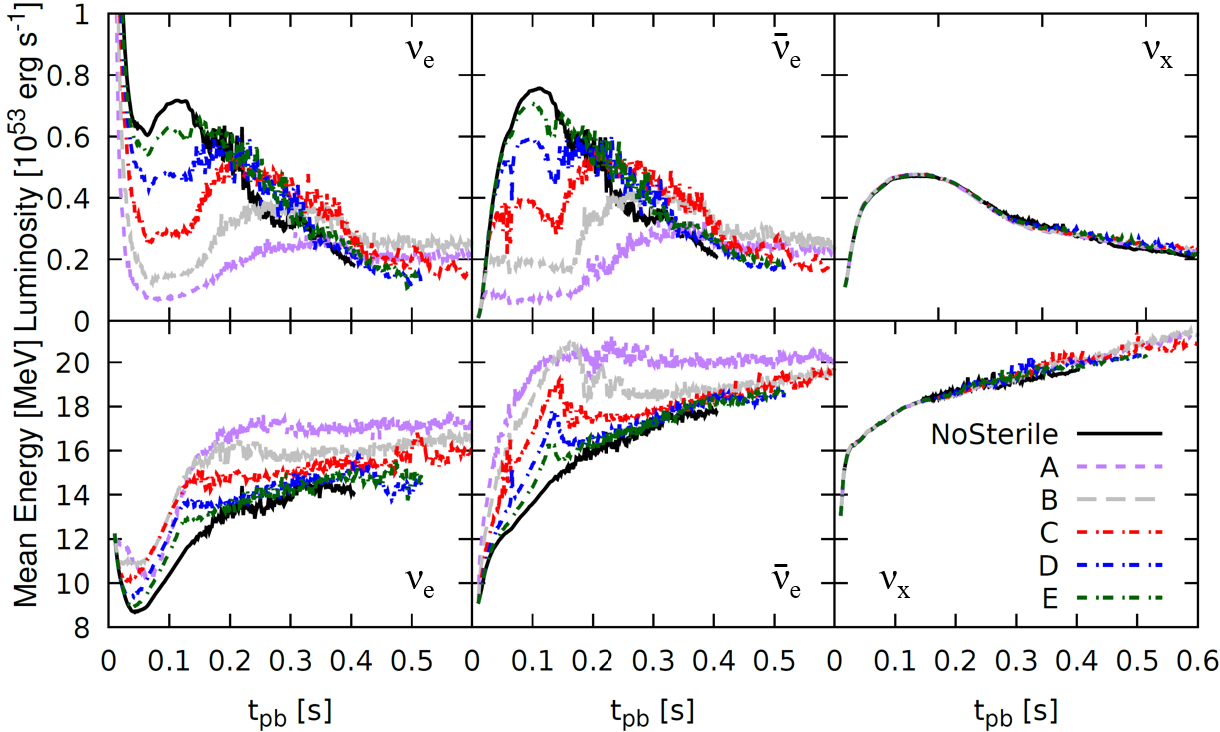}
  \caption{The active neutrino luminosity and the mean energy measured at $r=500$\,km as a function of $t_\mathrm{pb}$.}
  \label{neu}
 \end{figure*}

The hindrance of the shock revival and the reduction in the explosion energy are caused by the oscillation between $\nu_e (\bar{\nu}_e)$ and $\nu_s (\bar{\nu}_s)$. Fig.~\ref{r_res} shows the IR and OR radii for Model \texttt{C}. One can find that the IR always appears behind the stalled shock wave. This suggests that the IR can affect the neutrino heating in the gain region. On the other hand, the OR, which is not considered in our simulations, appears in the stellar envelope at $r\gtrsim10^3$\,km. It can also be seen that the angular dependence of the IR radius is within 30\,km. The OR radius does not depend on the zenith angle before the shock passage, but after that the OR depends on the angle because the matter profile is asymmetric in the post-shock region.

Figure \ref{P1} shows the $\nu_e$ survival probability $P_\mathrm{ee}(E)=1-P_\mathrm{es}(E)$ for the IR as a function of the neutrino energy $E$. One can find that, in our most extreme case, namely Model \texttt{A}, the $\nu_e$ flux is reduced by $\sim40$\,\% at $E=10$\,MeV. This reduction suppresses the neutrino heating rate in the gain region and hinder the shock revival, as shown in Fig.~\ref{rsh_2D}. We can also see that, if we compare $P_\mathrm{ee}(E)$ for each model,  their order follows that of $\delta m_\mathrm{s}^2\sin 2\theta$ shown in Table 1. This is because, as shown in Eq.~\eqref{LZ}, $P_\mathrm{es}(E)$ can be written as a function of $\delta m_\mathrm{s}^2\sin 2\theta$. When $\delta m_\mathrm{s}^2\sin 2\theta$ is larger, the resonance becomes more adiabatic and  $P_\mathrm{es}(E)$ becomes larger. It is then expected that the effect of sterile neutrinos on supernova dynamics is determined only by the combination of $\delta m_\mathrm{s}^2\sin 2\theta$. As seen in Fig.~\ref{Eexp}, the explosion energy at the end of the simulations follows the order of  $\delta m_\mathrm{s}^2\sin 2\theta$ for each model.

Figure \ref{Q} shows the neutrino heating rate in the gain region for Models \texttt{NoSterile}, \texttt{A}, and \texttt{C}. The rate is evaluated at $t_\mathrm{pb}=0.1$\,s, and at this moment the IR is located at $r\approx95.8$\,km for Model \texttt{A} and $r\approx99.7$\,km for Model \texttt{C}. In both of Models \texttt{A} and \texttt{C}, the IR is located in the gain region, and one can find that the heating rate is reduced at $r=r_\mathrm{IR}$ because of the active-sterile oscillation. In Model \texttt{A}, the reduction in the heating rate is more significant, and the net effect turns into cooling at $r\approx r_\mathrm{IR}$. This is because $\delta m_\mathrm{s}^2\sin 2\theta$ is larger and thus the electron (anti)neutrino survival probability is smaller, as shown in Fig.~\ref{P1}. 

Figure \ref{neu} shows the active neutrino luminosity and the mean energy measured at $r=500$\,km. One can see that, in the first $\mathcal{O}(0.1)$\,s after the core bounce, the electron (anti)neutrino luminosity $L_\nu$ is significantly reduced and its mean energy $\langle\epsilon_\nu\rangle$ is enhanced by sterile neutrinos. For example, in Model \texttt{A}, the electron neutrino luminosity is reduced by $\sim89$\% and its mean energy is enhanced by $\sim17$\% at $t_{\mathrm{pb}}=0.1$\,s. Since the reduction in $L_\nu$ hinders neutrino heating and the enhancement in $\langle\epsilon_\nu\rangle$ helps it, the effect on explosion dynamics is nontrivial. However, sterile neutrinos hinder supernova explosion in total because the neutrino heating rate is proportional to $L_\nu\langle\epsilon_\nu\rangle^2$. In a later phase, the electron (anti)neutrino flux in the models with sterile neutrinos exceeds the one in Model \texttt{NoSterile}. This is because the mass accretion rate is higher in Models \texttt{A}-\texttt{E}.

In Table 1, the ejected nickel mass $M_\mathrm{Ni}$ and the proto-neutron star mass $M_\mathrm{PNS}$ are shown. One can find that sterile neutrinos decrease $M_\mathrm{Ni}$ and increase $M_\mathrm{PNS}$, and they are monotonous functions of $\delta m_\mathrm{s}^2\sin 2\theta$. The reduction in the nickel mass is attributed to the reduced ejecta mass and the reduced neutrino heating which leads to lower entropy in the core. The proto-neutron star mass is enhanced because the mass accretion lasts until a later time if sterile neutrinos are considered.

The results presented here are significantly different from  previous studies \cite{2018PhRvD..98j3010R,2024PhRvD.110b3031M} on $\mathcal{O}(100)$\,MeV-mass sterile neutrinos. In the supernova models developed in the previous studies, the explosion became more energetic because of the $\nu_\mathrm{s}$ decay into photons and active neutrinos. Our results are opposite to theirs because our sterile neutrinos are stable and do not contribute to heating.

\subsection{Condition for Successful Explosion}\label{sec:co-su-exp}

The stalled shock wave found  in Models \texttt{A} and \texttt{B} leads to a condition on sterile neutrinos for successful supernova explosions. Since we adopted the progenitor model \cite{2018MNRAS.473L.101U} which is fine-tuned to reproduce the SN~1987A progenitor, our simulations are expected to reproduce SN~1987A explosion. The failed explosion in Models \texttt{A} and \texttt{B} indicates that SN~1987A explosion cannot be explained if $\delta m_\mathrm{s}^2\sin 2\theta\gtrsim0.45$\,eV$^2$. This argument provides an allowed region for  the mixing angle 
\begin{eqnarray}
    \sin 2\theta\lesssim0.45\,\textrm{eV}^2/\delta m_\mathrm{s}^2\label{constraint}
\end{eqnarray}
on the basis of SN~1987A explodability.

The condition~\eqref{constraint} is applicable only for eV-mass sterile neutrinos. In our simulations, $r_\mathrm{IR}$ is not sensitive to $\delta m_\mathrm{s}^2$ and $\sin^2 2\theta$ because 
\begin{eqnarray}
       \frac{\delta m_\mathrm{s}^2}{2E}\cos2\theta\ll G_\mathrm{F}n_\mathrm{b}\label{limit}
\end{eqnarray}
is satisfied. If this condition breaks down, $r_\mathrm{IR}$ depends on the sterile neutrino parameters and the models will not be explained as a function of $\delta m_\mathrm{s}^2\sin2\theta$. Plugging $\rho\approx10^{10}$\,g cm$^{-3}$ and $E\approx10$\,MeV into Eq.~\eqref{limit}, it turns out that Eq.~\eqref{constraint} is valid only when $\delta m_\mathrm{s}^2\ll 100$\,eV$^2$. 

Since the failed explosions observed for Models \texttt{A} and \texttt{B} are not compatible with SN~1987A explosion, Eq.~(\ref{constraint}) could be regarded as a new upper limit on the mixing angle. However, current numerical simulations for a collapsing star still have uncertainties including stellar evolution, the neutrino transport, and neutrino oscillations. In order to establish a new constraint based on supernova explodability, it is necessary to perform systematic investigation on uncertainties in Eq.~\eqref{constraint}, which is beyond the scope of this study.

\subsection{Neutrino Signals}

  \begin{figure}
  \centering
  \includegraphics[width=8.5cm]{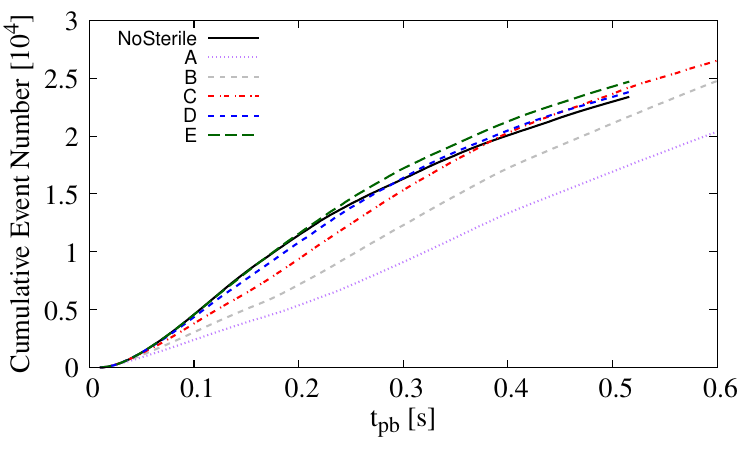}
  \includegraphics[width=8.5cm]{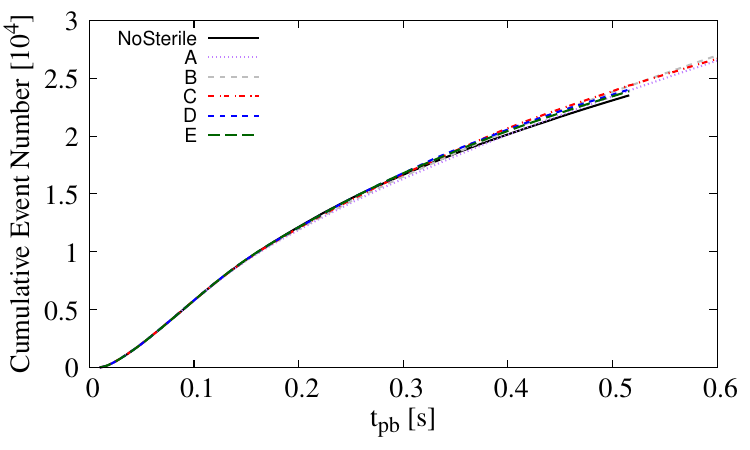}
  \caption{The cumulative $\bar{\nu}_e$ event number from a supernova event at $D=10$\,kpc away measured by Hyper-Kamiokande. The top panel assumes the NH and the bottom panel assumes the IH.}
  \label{L}
 \end{figure}

   \begin{figure}
  \centering
  \includegraphics[width=8.5cm]{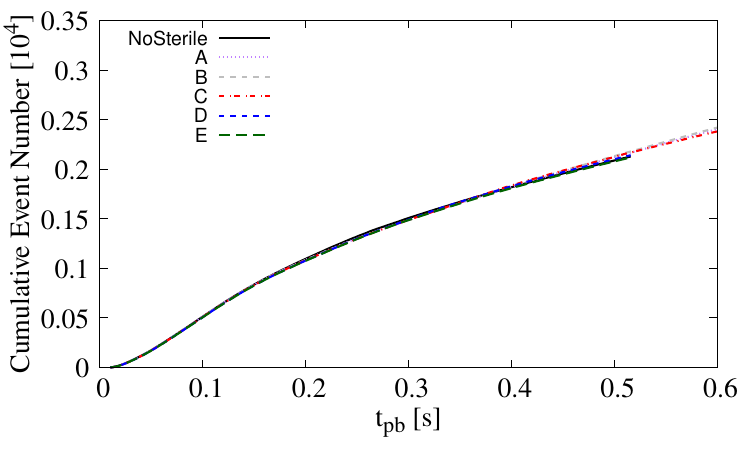}
  \includegraphics[width=8.5cm]{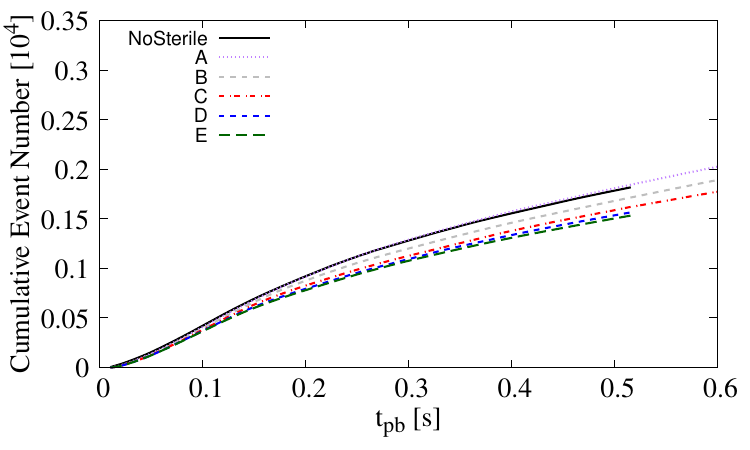}
  \caption{The cumulative $\nu_e$ neutrino event number from a supernova event at $D=10$\,kpc away measured by DUNE. The top panel assumes the NH and the bottom panel assumes the IH.}
  \label{Ar}
 \end{figure}

The oscillation between active and sterile neutrinos affects neutrino signals from a nearby supernova event. Electron (anti)neutrinos produced on the neutrinosphere first arrive at the IR. Since the IR is nonadiabatic, only a part of electron (anti)neutrinos become sterile (anti)neutrinos and the rest remains to be the original flavor. In the case of the neutrino sector, the neutrinos  arrive at the adiabatic OR, which causes the flavor swap. In the case of the antineutrino sector, however, the OR does not appear. Instead, after the IR, the antineutrinos encounter the H- and L-resonances \cite{2000PhRvD..62c3007D}, which are irrelevant to the sterile flavor.

The electron antineutrino flux $F(\bar{\nu}_e)$ that arrives on Earth is given as $F(\bar{\nu}_e)=pF_0(\bar{\nu}_e)+(1-p)F_0(\bar{\nu}_X)$ up to the geometrical factor, where $F_0$ is the flux of each flavor which has just passed through the IR. The conversion probability is given as $p=\cos^2\theta_{12}\cos^2\theta_{13}\approx0.682$ for the normal mass hierarchy (NH) and $p=\sin^2\theta_{13}\approx0.0220$ for the inverted hierarchy (IH) \citep{2024PhRvD.110c0001N}.  The flux $F_0$ is evaluated in the simulations, as shown in Fig.~\ref{neu}.

Figure \ref{L} shows the cumulative event number of electron antineutrinos detected by Hyper-Kamiokande \cite{2018arXiv180504163H,2021ApJ...916...15A}, which is a future water-Cherenkov detector. We assume a supernova event at $D=10$\,kpc away. The event rate per unit time is evaluated as
 \begin{eqnarray}
     \frac{dN_\nu}{dt}=N_\mathrm{tar}\int_{E_\mathrm{th}}^\infty F(E)\sigma(E) dE,
\end{eqnarray}
where $N_\mathrm{tar}$ is the number of the target protons, the $\sigma(E)$ is the  inverse $\beta$-decay cross sections \cite{2002RvMP...74..297B}, and $E_\mathrm{th}=8.3$\,MeV is the threshold energy \cite{2017ApJ...848...48K}. The figure shows that, in the NH case, the neutrino event number for Model \texttt{A}-\texttt{E} is smaller than the one in Model \texttt{NoSterile} in the first $\mathcal{O}(0.1)$\,s. This is because of the reduced $\bar{\nu}_e$ luminosity shown in Fig.~\ref{neu}. In Models \texttt{C}, \texttt{D}, and \texttt{E}, the event number becomes slightly larger than the one in Model \texttt{NoSterile} in a later phase because of the higher mass accretion rate. In Models \texttt{A} and \texttt{B}, the event number is smaller than the one for the model without sterile neutrinos until the end of the simulations. On the other hand, in the IH case, the neutrino event number is not affected by sterile neutrinos, because the contribution of heavy lepton  neutrinos is dominant and the effect of sterile neutrinos on the $\nu_X$ luminosity is small, as shown in Fig.~\ref{neu}. 

In the case of electron neutrinos, the flux $F(\nu_e)$ on Earth is given as $F(\nu_e)=pF_0(\nu_s)+(1-p)F_0(\nu_X)$. In this case $F_0(\nu_s)$ contributes to the $\nu_e$ flux because the flavor is swapped between $\nu_e$ and $\nu_s$ at the OR. The conversion probability is given as $p=\sin^2\theta_{13}\approx0.0220$ for NH and $p=\sin^2\theta_{12}\cos^2\theta_{13}\approx0.296$ for IH.

Electron neutrinos from a nearby supernova could be detected by a liquid scintillator such as the Deep Underground Neutrino Experiment (DUNE) \cite{2020JInst..15.8008A}. DUNE is a future liquid argon scintillator, which could detect $\nu_e$ through the charged-current reaction on $^{40}$Ar. In Fig.~\ref{Ar}, we estimate the $\nu_e$ event number from a supernova event at $D=10$\,kpc detected by DUNE, adopting the cross section that originates from Ref.~\cite{2003JPhG...29.2569K} and is tabulated in \texttt{SNOwGLoBES}\footnote{\protect\url{https://github.com/SNOwGLoBES/snowglobes}}. In the NH case, the result does not depend on the sterile neutrino parameters because the signal originates from heavy-lepton neutrinos. On the other hand, in the IH case, sterile neutrinos affect the signal and the deviation from Model \texttt{NoSterile} is larger when the mixing angle is smaller. This counterintuitive behavior is explained as follows. When the mixing angle is small, the $\nu_s$ flux produced at the IR is small. However, since the OR is adiabatic in our parameter region, the small $\nu_s$ flux leads to a small $\nu_e$ flux. This result implies that observations of $\nu_e$ from a nearby supernova event would provide a stringent constraint on the sterile neutrino parameters, as pointed out in Ref.~\cite{2020JCAP...10..038T}. 
\subsection{Gravitational Waves}

  \begin{figure}
  \centering
  \includegraphics[width=8.5cm]{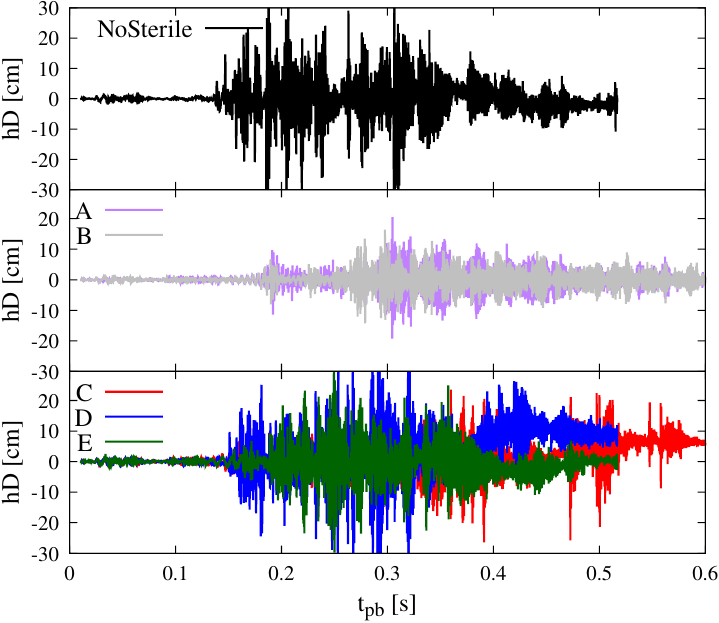}
  \caption{The GW plus-mode waveform $h_+D$ from a supernova event at $D=10$\,kpc. The observer is assumed to be located at the equatorial direction of the star.}
  \label{GW_waveform}
 \end{figure}
   \begin{figure}
  \centering
  \includegraphics[width=8.5cm]{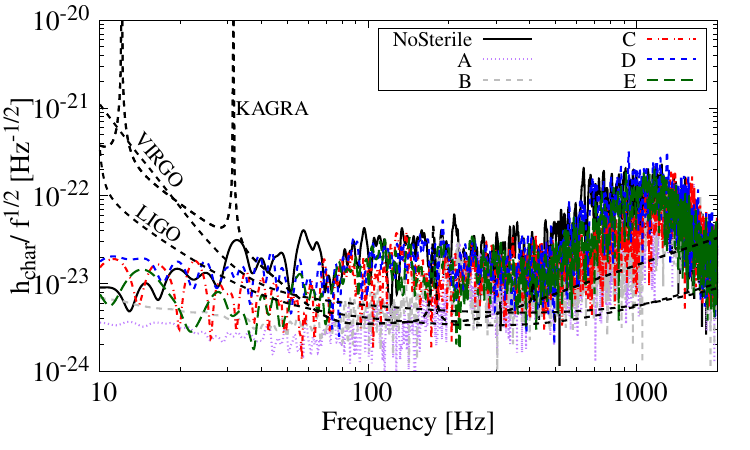}
  \caption{The GW spectrum and the sensitivity of ground-based detectors, namely LIGO, VIRGO, and KAGRA. The distance to the supernova event is assumed to be 10\,kpc.}
  \label{GW_spec}
 \end{figure}

Galactic core-collapse supernova events are a possible target for GW astronomy. Since GWs keep information inside the stellar core, they could be a unique probe for physics in the supernova engine. With our axisymmetric models, we can predict the GW plus-mode emission from a nearby event. The GW strain can be estimated as \cite{2009ApJ...707.1173M}
\begin{eqnarray}
    h_+=\frac{3}{2}\frac{G}{Dc^4}\sin^4\alpha\frac{d^2}{dt^2}\ibar_{zz},
\end{eqnarray}
where $G$ is the gravitational constant, $c$ is the speed of light, $\alpha$ is the angle between the line of sight and the polar axis, and  $\ibar_{zz}$ is the reduced quadruple moment.

Figure \ref{GW_waveform} shows the GW waveform $h_+D$ with $D=8.5$\,kpc, assuming $\alpha=\pi/2$. For all models, the strong GW emission starts at $t_\mathrm{pb}\sim0.15$--0.2\,s. At this phase, non-radial motion similar to the standing accretion shock instability sets in. One can also see that the GW amplitude for Models \texttt{A} and \texttt{B}, which do not show the shock revival, is smaller than that in the other models because the motion of the matter is more spherically symmetric.

Figure \ref{GW_spec} shows the  strain \cite{PhysRevD.57.4535}
\begin{eqnarray}
h_\mathrm{char}=\sqrt{\frac{2G}{\pi c^3D^2}\frac{dE_\mathrm{GW}}{df}},
\end{eqnarray}
as a function of the GW frequency $f$. Here $dE_\mathrm{GW}/df$ is the GW spectral energy density. Sensitivity curves for ground-based GW detectors LIGO \citep{2015CQGra..32g4001L}, VIRGO \citep{2015CQGra..32b4001A}, and KAGRA \citep{2021PTEP.2021eA101A} are also plotted in the figure. One can find that all of our models lead to detectable signals if the supernova event appears at the Galactic center.

Figure \ref{GW_spec} indicates that all of the models with and without sterile neutrinos show a similar peak at $f\sim1$\,kHz. This high-frequency GW is emitted from the proto-neutron star. The effect of the active-sterile oscillations on this peak is minor because sterile neutrinos are produced outside of the proto-neutron star and their effect on $M_\mathrm{PNS}$ is less than 10\%. In Fig.~\ref{GW_waveform}, one can also find that the GW waveform shows the ``memory" effect~\cite{2009ApJ...707.1173M} for Models \texttt{C} and \texttt{D}, whose explosion morphology is strongly prolate.


\section{Summary and Discussion}\label{sec:sum-dis}

In this work, we performed two-dimensional simulations for a core-collapsing star coupled with the oscillations between $\nu_e$ and $\nu_s$. It was found that the oscillations reduce the neutrino heating rate and hinder supernova explosions. We obtained a condition to obtain a successful SN~1987A explosion, $\sin 2\theta\lesssim0.45\,\textrm{eV}^2/\delta m_\mathrm{s}^2$, for light sterile neutrinos with $\delta m_\mathrm{s}^2\ll100$\,eV$^2$  and predicted multi-messenger signals from a Galactic supernova event in the future. In principle, the condition for the successful explosion could be regarded as a constraint on sterile neutrinos. However, the constraint cannot be established by our results because our simulations are subject to model uncertainties. Instead, this study provides the formulation to implement the active-sterile oscillations in simulations and a prospect for the new constraint.  Future improvement on supernova simulations would reduce the uncertainties and provide a constraint on the sterile neutrino mass and the mixing angle.

Our simulations adopt artificial axisymmetric geometry to reduce computational resources. In general, two-dimensional models explode more easily \cite[e.g.,][]{2018ApJ...852...28S} if other inputs including progenitor models and equations of state are fixed. Our constraint based on explodability is conservative in this sense and systematic three-dimensional simulations would lead to a tighter constraint.  

If we adopt the condition in Eq.~\eqref{constraint} as a constraint on sterile neutrinos, it  excludes a part of the parameter region preferred by $\nu_e$ disappearance experiments, as shown in Fig.~\ref{param}. However, there is still a room for sterile neutrinos favored by the experiments. In order to explore the possibility of solving the RAA with sterile neutrinos, further experimental and astrophysical studies are desirable. In particular, the neutronization burst from a future nearby supernova event would provide a tight constraint that can exclude the sterile neutrino solution for the RAA, as pointed out in Ref.~\cite{2020JCAP...10..038T}. 

We found that the active-sterile oscillation can reduce the neutrino event number from a nearby supernova event observed by terrestrial detectors. The $\bar{\nu}_e$ flux, which can be detected by water-Cherenkov telescopes such as Super- and Hyper-Kamiokande, decreases as a function of the $\nu_e$-$\nu_s$ mixing angle because antineutrinos experience only the IR. On the other hand, the $\nu_e$ flux, which can be detected by  scintillators such as DUNE, behaves in the opposite way as a function of the mixing angle because neutrinos experience both of the IR and the OR. As long as the OR is adiabatic, the $\nu_e$ flux will be significantly reduced compared with the one for the model without sterile neutrinos \cite{2020JCAP...10..038T}. This implies that $\nu_e$ signals from a nearby supernova event will be a key to investigate the nature of sterile neutrinos. 

Previous works \cite{2014PhRvD..89f1303W,2015ApJ...808..188P,2019ApJ...880...81X,2020ApJ...894...99K} have pointed out that light sterile neutrinos can significantly affect supernova nucleosynthesis as well. In particular, they report that $Y_e$ in the ejecta and the neutrino-driven wind is lowered because of the reduced $\nu_e$ flux. In our simulations, however, the reduction in $Y_e$ was not obvious. This difference may be attributed to the interplay between the reduced $\nu_e$ flux and hydrodynamics. The explosion energy and the ejecta mass are different in each model, and in addition, the convective motion in our two-dimensional models makes it difficult to simply understand the $Y_e$ distribution in the ejecta because the neutrino exposure time is different for each mass element. Also, our treatment on the feedback of the active-sterile oscillations on trapped neutrinos is different from the previous works. This may affect the ratio between the $\nu_e$ and $\bar{\nu}_e$ fluxes, and consequently, $Y_e$. Since supernovae drive the galactic chemical evolution, it is worthwhile to investigate the effects of sterile neutrinos on nucleosynthesis  by performing long-term multi-dimensional hydrodynamic simulations coupled with the neutrino oscillations. 

It has been also pointed out that sterile neutrinos can appear in neutron-star mergers \cite{2022PhRvD.106l3030S}, which is one of the production sites of heavy elements in the Universe. Impacts of sterile neutrinos on dynamics of neutron-star mergers are not fully investigated, and hydrodynamical simulations coupled with the oscillations between active and sterile neutrinos are needed. 

Our simulations performed in this study consider the MSW conversion between active and sterile neutrinos. However, the sterile neutrino flux could also be affected by collective oscillations induced by the neutrino self-interaction  \cite{2012JCAP...01..013T,2019ApJ...880...81X}. In the context of the active-active oscillations, the stability analysis for the quantum kinetic equation (QKE) has been performed to reveal that the fast flavor instability \cite{2019PhRvD.100d3004A,2019ApJ...886..139N,2021PhRvD.103f3033A,2021PhRvD.104h3025N,2024PhRvD.109b3012A} and the collisional flavor instability \cite{2023PhRvD.108l3024L,2024PhRvD.109b3012A} can take place even in the optically-thick region. In addition, the QKE has been solved in supernova backgrounds to follow the non-linear growth of the fast flavor instability \cite{Nagakura2023,Xiong2024} and the collective flavor instability \cite{2023PhRvD.107h3016X}. In order to investigate the impact of the collective oscillations, it is worthwhile to analyze the QKE with sterile neutrinos, following these studies. 

In this work, we focused on eV-mass sterile neutrinos. It would be also interesting to develop stellar core-collpse simulations coupled with keV-mass sterile neutrinos \cite[e.g.][]{2014PhRvD..90j3007W,2019JCAP...12..019S,2022PhRvD.106a5017S,2023PhRvD.108f3025R,2024PhRvD.110d3007R}, which are a candidate for dark matter \cite{1994PhRvL..72...17D,1999PhRvL..82.2832S}. When the sterile neutrino mass becomes much larger than the eV-scale, the IR becomes located at a deeper radius. As a result, the IR can appear at a smaller radius, where the active neutrino distribution is more isotropic. It is desirable to develop a neutrino transfer scheme that can treat such heavy sterile neutrinos to extend our supernova constraints to the heavier region.

This study marks a steady step toward understanding supernova neutrino oscillations and their impact including contributions from the non-active sector~\cite{Sasaki2021,Sasaki2023,Chauhan2024}. Given the crucial role of neutrino heating in the explosion mechanism, as well as the significant potential of using neutrino signals from a future Galactic supernova to unveil key aspects of supernova physics and neutrino properties~\cite{Suwa_2011,Nakazato2022,Suwa2019,Harada2023,Suwa2025}. To achieve a comprehensive understanding of the effects of neutrino oscillations in supernovae~\cite{Nagakura2023,Xiong2024,Ehring2023,Mori2025arXiv,Wang2025arXiv,Duan2010,Mirizzi2016,Tamborra2021,Capozzi2022review,Richers2022review,Volpe2023review} and the complex physics involved~\cite{Kotake06,Janka12,Burrows21,Yamada2024}, these studies must consider all the discussed factors across a wide range of models with different progenitors~\cite{O'connor2011,Ebinger2020,Couch2020,Warren2020,Boccioli2023, Zapartas2021,Sasaki2024,2024ApJ...964L..16B,2025MNRAS.536..280N}.

\begin{acknowledgments}
KM is thankful to Kei Kotake and Shunsaku Horiuchi for useful comments. Numerical computations were  carried out on Cray XC50 and XD2000 at Center for Computational Astrophysics, National Astronomical Observatory of Japan. This work is supported by JSPS KAKENHI Grant Numbers JP23KF0289, JP24H01825,  JP24K07027, JP23KJ2147,  JP23K03468, JP23K13107, JP23K20848, JP23K22494, JP23K25895, JP23K03400, JP24K00631, and World Premier International Research Center Initiative (WPI Initiative), MEXT, Japan.
\end{acknowledgments}

\appendix
\section{Collapsing Phase}\label{sec:copha}
In our two-dimensional simulations, we switched on the active-sterile oscillations at $t_\mathrm{pb}=10\,$ms. This treatment can be justified because the IR becomes highly non-adiabatic in the collapsing phase, as shown below.

In the collapsing phase, the matter in the core consists of not only nucleons but also nuclei. In this case, the matter potential shown in Eq.~\eqref{Veff} should be modified. In general, the matter potential is written as 
\begin{eqnarray}
V_\mathrm{eff}=V_\mathrm{CC}+V_\mathrm{NC},
\end{eqnarray}
where $V_\mathrm{CC}$ is the charged-current potential and $V_\mathrm{NC}$ is the neutral-current potential. Although the neutral-current component does not affect the MSW resonance for active neutrinos, it does matter for sterile neutrinos, which are not involved with the neutral-current interaction. For $\nu_e$, the charged-current component is given as 
\begin{eqnarray}
V_\mathrm{CC}=\sqrt{2}G_\mathrm{F}n_e,
\end{eqnarray}
where $n_e$ is the electron number density. On the other hand, the neutral-current component for any active neutrino $\nu_\alpha$ and a fermion $f$ is given as 
\begin{eqnarray}
V_\mathrm{NC}^f=\sqrt{2}G_\mathrm{F}n_fg_V^f,
\end{eqnarray}
where $N_f$ is the number density of the fermion $f$. The coefficients $g_V^f$ for electrons, protons, and neutrinos are given as \cite[e.g.,][]{10.1093/acprof:oso/9780198508717.001.0001}
\begin{eqnarray}
g_V^e&=&-\frac{1}{2}+2\sin^2\theta_\mathrm{W},\\
g_V^p&=&+\frac{1}{2}-2\sin^2\theta_\mathrm{W},\\
g_V^n&=&-\frac{1}{2},
\end{eqnarray}
where $\theta_\mathrm{W}$ is the Weinberg angle. The coefficient $g_V^{(Z,\,N)}$ for nuclei with the atomic number $Z$ and the neutron number $N$ can be estimated as
 \begin{eqnarray}
g_V^{(Z,\,N)}=Zg_V^p+Ng_V^n=\left(\frac{1}{2}-2\sin^2\theta_\mathrm{W}\right)Z-\frac{N}{2}.
\end{eqnarray}
The total matter potential is then given as
\begin{align}
V_\mathrm{eff}=\sqrt{2}G_\mathrm{F}\left(\left(\frac{1}{2}+2\sin^2\theta_\mathrm{W}\right)n_e\right.\nonumber \\
\left.+\sum_{(Z,\,N)}\left(\left(\frac{1}{2}-2\sin^2\theta_\mathrm{W}\right)Z-\frac{N}{2}\right)n_{(Z,\,N)}\right), \label{V_general}
\end{align}
where $n_{(Z,\,N)}$ is the number density of the nucleus. Using the charge neutrality $\sum_{(Z,\,N)}Zn_{(Z,\,N)}-n_e=0$
and the definition for the baryon number density
$\sum_{(Z,\,N)}(Z+N)n_{(Z,\,N)}=n_\mathrm{b}$, we obtain
\begin{eqnarray}
 V_\mathrm{eff}=\frac{3\sqrt{2}}{2}G_\mathrm{F}n_\mathrm{b}\left(Y_e-\frac{1}{3}\right),
\end{eqnarray}
which coincides with the right hind side of Eq.~\eqref{Veff}. This implies that the IR appears at the isosurface of $Y_e\approx1/3$ even under the existence of nuclei.

\begin{figure}
  \centering
  \includegraphics[width=8.5cm]{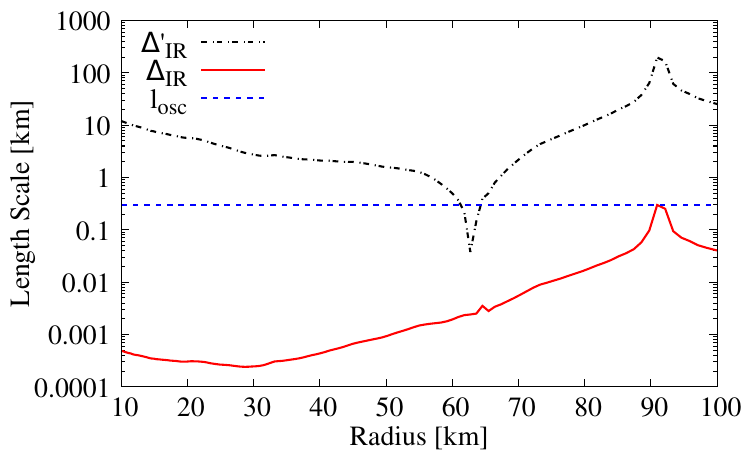}
  \caption{The length scales related with the active-sterile oscillations at $t_\mathrm{pb}=-2$\,ms for Model \texttt{1D-C}. The neutrino energy is fixed to $E=10\,$ MeV to estimate $\Delta_\mathrm{IR}$.}
  \label{scale}
 \end{figure}
 
In order to investigate the active-sterile oscillations in the collapsing phase, we perform a one-dimensional simulation with  sterile neutrinos. Fig.~\ref{scale} shows length scales at $t_\mathrm{pb}=-2$\,ms related with the active-sterile oscillations for Model \texttt{1D-C}.
Here the IR width is defined as
\begin{eqnarray}
\Delta_\mathrm{IR}&=&\left|\frac{dV_\mathrm{eff}}{dr}\right|^{-1}\frac{\delta m_\mathrm{s}^2}{2E}\sin2\theta,\label{d1}
\end{eqnarray}
and the oscillation length is defined as $l_\mathrm{osc}=2\pi E/\delta m_\mathrm{s}^2\sin2\theta$. 
Eq.~\eqref{d1} is adopted to estimate the IR width and the conversion probability in our simulations.

Another mathematically possible expression for the IR width is
\begin{eqnarray}
\Delta'_\mathrm{IR}&=&\left|\frac{\frac{dV_\mathrm{eff}}{dr}}{V_\mathrm{eff}}\right|^{-1}\tan2\theta.
\end{eqnarray}
The two expressions, $\Delta_\mathrm{IR}$ and $\Delta'_\mathrm{IR}$, are equivalent to each other on resonance, considering the resonance condition in Eq.~(\ref{res}).
In Model \texttt{1D-C}, the IR starts appearing at $t_\mathrm{pb}=-8$\,ms, when $Y_e$ becomes smaller than 1/3 in the core. At $t_\mathrm{pb}=-2$\,ms, the IR is located at $r\approx63$\,km. One can find that $\Delta'_\mathrm{IR}$ has a sudden dip at $r\approx r_\mathrm{IR}$ because $V_\mathrm{eff}$ at the IR is close to zero and thus the $V_\mathrm{eff}$ scale hight $h=|(dV_\mathrm{eff}/dr)/V_\mathrm{eff})|^{-1}$ becomes short. Because of the resonance condition in Eq.~(\ref{res}), $\Delta_\mathrm{IR}$ should be equal to $\Delta'_\mathrm{IR}$ on resonance. However, in Fig.~\ref{scale}, they do not coincide with each other at $r=r_\mathrm{IR}$  because the spatial resolution is not enough to resolve $\Delta'_\mathrm{IR}$ at $r\approx r_\mathrm{IR}$. This is why one should use Eq.~\eqref{d1} in simulations to estimate the IR width correctly.

We can also see that $\Delta_\mathrm{IR}$ is shorter than $l_\mathrm{osc}$ at the IR. This implies that the IR is non-adiabatic  because the $V_\mathrm{eff}$ profile is steep at this phase. Because the resonance is non-adiabatic, the $\nu_e$ conversion probability is as low as $P_\mathrm{es}(E)=0.03$ for $E=10$\,MeV and thus the effect of the oscillations on the collapsing phase is not considered in our two-dimensional models.

\bibliography{ref.bib,refarXiv}
\end{document}